# Sound Wave Manipulation Based on Valley Acoustic Interferometers

*Wei Zhao, Jia-He Chen, Shu-Guang Cheng, Yong Mao, Xiaojun Zhang, Zhi Tao, Hua Jiang[*],*

*and Zhi Hong Hang[*]*

W. Zhao, J.-H. Chen, Y. Mao, H. Jiang, Z. H. Hang
School of Physical Science and Technology & Collaborative Innovation Center of Suzhou Nano Science and Technology
Soochow University
Suzhou 215006, China
E-mail: jianghuaphy@suda.edu.cn; zhhang@suda.edu.cn

W. Zhao
School of Electronic and Information Engineering
Suzhou Vocational University
Suzhou 215100, China

S.-G. Cheng
Department of Physics
Northwest University
Xi'an 710069, China

X. Zhang, Z. Tao
School of Optoelectronic Science and Engineering
Soochow University
Suzhou 215006, China

H. Jiang, Z. H. Hang
Institute for Advanced Study
Soochow University
Suzhou 215006, China

H. Jiang, Z. H. Hang
Jiangsu Key Laboratory of Frontier Material Physics and Devices
Soochow University
Suzhou 215006, China

**Abstract:** Topological acoustics provides new opportunities for materials with unprecedented functions. In this work, we report a design of topological valley acoustic interferometers by Y-shaped valley sonic crystals. By tight-bounding calculation and experimental demonstration, we successfully tune the acoustic energy partition rate by configuring the channel. An analytical theory proposed to explain the transmission property matches well with experimental observations. An additional π Berry phase is verified to accumulate circling along the shape-independent topological valley acoustic interferometer, unique in the pseudospin half systems. Based on the spectral oscillation originating from the accumulated dynamic phase and π Berry phase, a simplified method to measure acoustic valley interface dispersion is explored, which overcomes the shortcomings of the traditional fast Fourier transform method and improves the measuring efficiency by simply analyzing the peaks and dips of the measured transmission spectrum. Moreover, an effective approach to tuning its transmissions, as well as the spectral line shapes proposed and realized by the local geometry design of the interferometer, exhibits strong tunability under an unchanged physical mechanism. Our work opens an avenue to design future acoustic devices with the function of sound wave manipulation based on the physical mechanism of interference and Berry phase.

## 1. Introduction

Valley pseudospin, labeling the degenerate energy extrema in momentum space, has sparked extensive interest in condensed matter physics due to its potential to serve as an additional degree of freedom to manipulate electrons besides charge and spin.[1-7] In recent years, topological valley interface states, featuring gapless dispersion and robust electron transport, have been observed around the domain walls with opposite valley Chern numbers by breaking inversion symmetry in graphene systems.[8-16] Based on the unique backscattering-immune characteristics of valley-momentum locking of topological valley interface states, various valley devices[12-17] including valley filters, valley valves, and valley splitters, have thus been proposed. An Aharanov-Bohm interferometer[18] composed of topological valley interface states has been proposed and utilized to periodically manipulate the valley current by applying a magnetic field, which paves the road for future applications of valleytronics.

Because of their similar band diagrams, sonic crystals (SCs) have been intensively studied,[19-37] especially to realize acoustic valley pseudospin.[22-37] Robust one-way acoustic wave propagation was observed in the domain walls constructed by SCs with inversion or mirror symmetry breaking. Benefiting from the flexibility of SCs, various acoustic devices, and effects such as directional acoustic antenna,[26] acoustic delay line,[25] acoustic switch,[31] topological refraction,[29,34] acoustic logic gates,[32,37] and topological rainbow concentrators[36] have been realized, which have potential applications in modern acoustic signal processing. The design and manipulation of topological domain properties allow multichannel topological energy partition devices[29,31-32,38-42] readily available. However, other than the construction of acoustic splitters with valley SCs, an effective way to arbitrarily manipulate the energy partition is needed. The role of the pseudospin degree of freedom plays in the energy partition device needs to be discussed.

In this work, we report topological valley acoustic interferometers constructed by Y-shaped aluminum valley SCs. By rotating valley SCs, we successfully tune the acoustic energy partition rate by reconfiguring the channel. The basic mechanism to achieve and tune the valley acoustic interferometer is discovered. The π Berry phase in graphene has been theoretically predicted[43] but it has only been very recently measured by observing the wavefront dislocation in Friedel oscillation.[44] In this work, for the very first time in classical wave system, we discover an extra π phase accumulates in addition to the dynamic phase accumulation along the circumference of the shape-independent topological domain. By the local geometry design of the interferometer, we can also effectively tune different properties of acoustic energy partition devices, such as line shapes and transmission rate where applications are thus intrigued.

## 2. Acoustic Energy Partition by Valley Sonic Crystals

As depicted in **Figure 1**a,b, the valley SC we design is composed of a hexagonal array (with the lattice constant of $a = 22$ mm) of Y-shaped aluminum scatterers in air. Each scatterer (with a height of $h = 8$ mm) consists of three legs with a length of $t = 6$ mm and a width of $w = 3$ mm, and the angle between two neighbor legs is 120°. Finite-element method (COMSOL Multiphysics) is used for the band diagram calculations and sound wave simulations throughout this work. The material parameters adopted are as follows: the density $\rho = 2700$ kg/m$^3$ and the sound velocity $c = 6300$ m/s for aluminum scatterers; $\rho = 1.21$ kg/m$^3$ and $c = 343$ m/s for air.

When the rotation angle $\chi$ is set as 0°, the $C_{3v}$ symmetry of the lattice is commensurate with Y-shaped scatterers, and a twofold degenerate linear dispersion, *aka* Dirac cone, appears at the corners of the first Brillouin zone (BZ). By rotating each scatterer at a rotation angle of $\chi = \pm 30°$, the mirror symmetry of SC is broken, and a band gap appears between 7183 Hz and 9117 Hz, as shown in Figure 1c. Though the acoustic band diagrams seem identical for $\chi = 30°$ and $\chi = -30°$, the characteristics of different topological phases between them can be found after comparing their energy flows of the intrinsic fields. For simplicity, here we only show the topological properties of the valley states at point $K$, and those at the inequivalent point $K'$ are readily available by time-reversal symmetry. As shown in the left panel of Figure 1c, for $\chi = 30°$, clockwise (anticlockwise) energy flow is clearly visualized in the insets at the lower (higher) frequency, and the corresponding acoustic valley state is donated as $K^{\downarrow(\uparrow)}$. As shown in the right panel of Figure 1c, for $\chi = -30°$, the corresponding band property with inverted vortex chirality can be obviously observed. We thus obtain the valley SCs needed.

By assembling two valley SCs belonging to different topological phases, valley interface states can thus be aroused at their domain walls. Figure 1d illustrates the calculated band diagrams of the supercells composed of two types of aluminum scatterers (pink for $\chi = -30°$ and cyan for $\chi = 30°$, as shown in the inset of Figure 1d). Along the different zigzag interfaces (denoted as orange/purple lines) between these valley SCs, there exist valley interface states, carrying different pseudospins, which originate from non-zero and opposite valley Chern numbers of valley SCs. According to the directions of the group velocities, the valley interface states are divided into the forward mode (pseudospin state $\varphi^+$, denoted as purple solid line) and backward mode (pseudospin state $\varphi^-$, denoted as orange solid line), which are in accord with the energy-flow directions (black arrows) shown in the insets in Figure 1c.

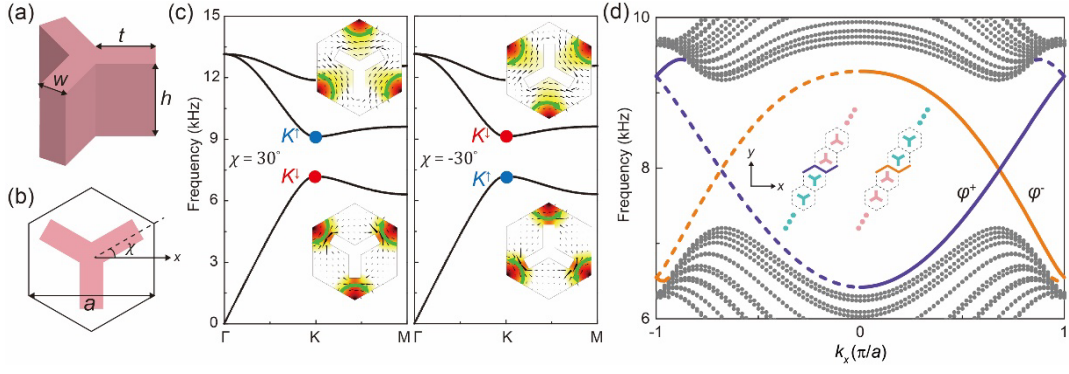

**Figure 1.** Topological valley SCs, topological band inversion, and topological protected valley interface states. a) 3D and b) 2D schematics of the SC composed of a Y-shaped aluminum block arranged in a hexagonal lattice. c) The bulk acoustic band diagrams with a rotation angle of $\chi = 30°$ (left panel) and -30° (right panel). The symbols $K^{\downarrow(\uparrow)}$ denote the valley states with clockwise (anticlockwise) energy flow as shown in the insets. d) Band diagram of valley interface states with different supercells and interfaces as shown in the inset. The mode with clockwise (anticlockwise) in-plane angular momentum is shown in the dashed (solid) line with colors to indicate the corresponding interface (denoted as the orange/purple line).

It has been predicted that the energy partition rates along different valley domain walls will be different when valley topological insulator domains are assembled at different intersection angles[16], and further experimentally verified in acoustics[42] and elastic waves.[38] We are going to elaborate that this is vital for our design of valley interferometers. Two sets of devices, namely Device 1 and Device 2, are prepared with different corresponding intersection angles $\theta_1$ and $\theta_2$, as shown in **Figure 2**a,c. The pink and cyan scatterers denote valley SCs with rotation angles $\chi = -30°$ and $\chi = 30°$, respectively. We first consider Device 1 with samples under the condition of $\theta_1+\theta_2 = 180°$ (as shown in Figure 2a, where $\theta_1$ and $\theta_2$ indicate the angles at the intersection points). A function generator (Tektronix AFG 3022C) is used to excite the speaker at Port 1 to allow sound wave incidence, and we measure sound transmission to Port 2, Port3, and Port 4 using a 1/4-inch microphone (BSWA MPA401) and a lock-in amplifier (Stanford Research SR830) at 8000 Hz for various $\theta_1$, as illustrated in Figure 2b. Two parallel plexiglass plates are used to clad these scatterers to guarantee a single mode of sound wave propagation and avoid leakage in the experiment. In the process of propagation, the emitting sound energy from Port 1 propagates forward and reaches the intersection point. Further propagation to Port 4 is impossible because $\varphi^+$ has to propagate backward along this domain wall.[32-33,39] Thus, the sound energy will detour to Port 2 and Port 3. It is seemingly counterintuitive that the sound energy partition to Port 3 is enhanced along a more detoured path with the increase of $\theta_1$. The backscattering-immune nature of valley interface states guarantees that sound propagation shall not be hindered by the sharp corner, and the stronger coupling to the domain wall leading to Port 3 at larger $\theta_1$ stimulates the corresponding energy partition. Though experimentally only discrete angles $\theta_1$ are used, we verify our experimental findings with tight-binding (TB) calculations with many more angles using the graphene-like system with fitted parameters. Consistent results are found in our experiments (symbols) and TB calculations (solid lines). In other words, the acoustic platform we develop here can be easily extended to the graphene-like system, and the results to be presented in this work are direct demonstrations of the corresponding condensed matter theories.

Moreover, it is also possible to tune the energy partition rate by reconfiguring the domain wall to Port 4 even though no sound propagation is allowed. As shown in Figure 2c, a configuration of Device 2 with $\theta_1 = 120°$ and variable $\theta_2$ is exhibited, which can be easily realized by simply tuning the rotation angles of several scatterers near the intersection region. As shown in Figure 2d, using numerical experiments we find that the transmission to Port 3 reduces and that to Port 2 rises with the increase of $\theta_2$. Similar consistency is also found between our numerical simulations and TB calculations. The results of the energy partition rate of valley SCs lay a solid foundation for further designing acoustic valley interferometers, which will be discussed in the following sections.

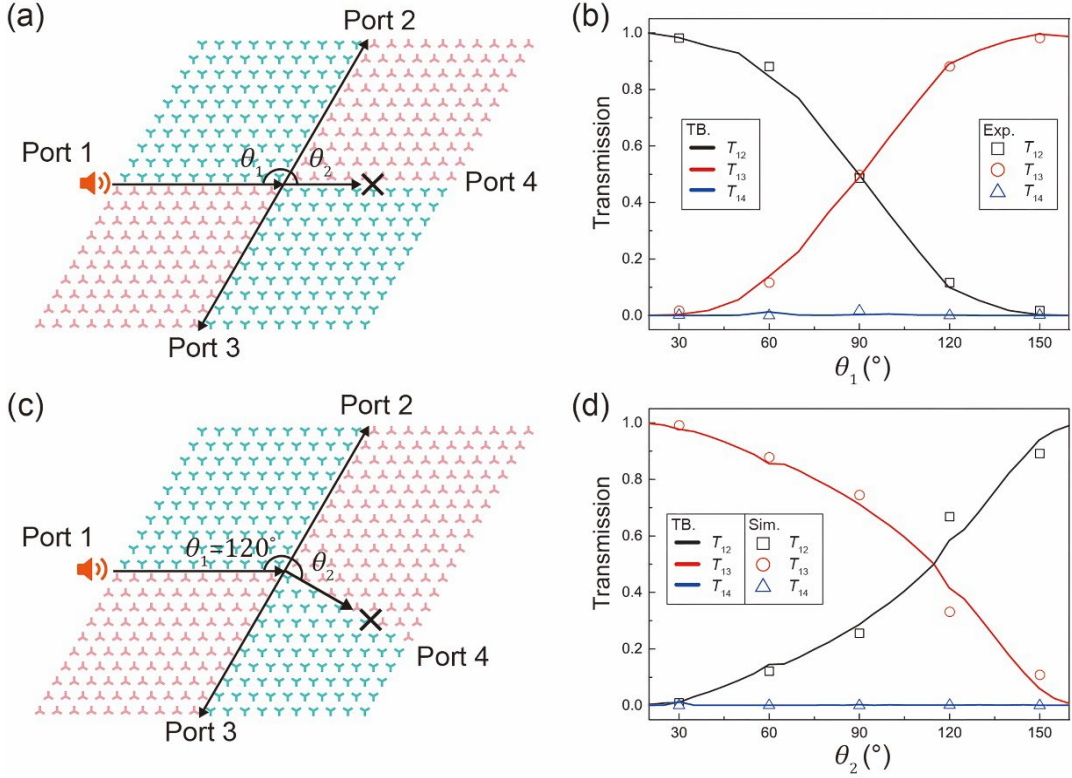

**Figure 2.** Acoustic energy partitions with different valley SC configurations. a) The schematic of Device 1, a valley energy splitter under the condition of $\theta_1+\theta_2 = 180°$. Sound wave is always incident from Port 1. b) The corresponding transmission spectra to Port 2 (black line for TB calculations and black open squares for experiments), Port 3 (red line for TB calculations and red open circles for experiments), and Port 4 (blue line for TB calculations and blue open triangles for experiments) at 8000 Hz for Device 1. c) The schematic of Device 2, a valley energy splitter under the condition of $\theta_1 = 120°$. d) The corresponding transmission spectra to Port 2 (black line for TB calculations and black open squares for simulations), Port 3 (red line for TB calculations and red open circles for simulations), and Port 4 (blue line for TB calculations and blue open triangles for simulations) at 8000 Hz for Device 2.

## 3. Mechanism Analysis of Valley Acoustic Interferometer

The study of topology physics starts from the unidirectional propagation of interface states between different topological domains, while intriguing effects aroused from finite-sized samples are discovered in classical wave systems.[45-57] We here also consider finite-sized experimental samples with firstly a rhombus-shaped topological acoustic interferometer of the size of $12a \times 12a$, as shown in **Figure 3**a. The aluminum scatterers with rotation angles $\chi$

= -30° and $\chi$ = 30° are adopted to form the domains with different topological phases. As shown in Figure 3a, the domain walls dividing different domains are donated as red dashed lines, which generate three small rhombus domains (one in the middle and two at the corners) with two intersections. The total circumference of a rhombus-shaped interferometer with four domain walls is $L = 48a$ as shown in Figure 3a. We start to discuss how sound wave shall propagate within such a topological interferometer. With the incidence from Port 1, the acoustic wave with $\varphi^+$ pseudospin is partitioned to Port 2 and the left domain wall of the interferometer at the first intersection. Note that the corresponding energy partition rate is fixed as we experimentally demonstrated in Figure 2. At the second intersection, the sound wave has no choice but to redistribute between Port 3 and the bottom domain wall of the interferometer at a ratio, which is also determined by the geometry at this intersection. No sound wave shall reach Port 4 due to a different $\varphi^-$ pseudospin allowed. Sound wave shall continue propagating without backscattering along the bottom, right, and top domain walls of the interferometer and finally reach the first intersection again. Redistribution between the left domain walls and Port 2 occurs again, and sound waves thus interfere inside. Obviously, transmission back to Port 1 is impossible, which means 100% of sound energy injected shall be distributed between Ports 2 and 3 based on the geometry of the interferometer.

By rotating several SCs, a smaller rhombus-shaped interferometer can be studied whose circumference is tuned to $L = 36a$, and its characteristics and comparison with $L = 48a$ are exhibited in Figure 3b. Obviously, similar spectral oscillation occurs, indicating their strong capability to manipulate sound wave distribution. Moreover, a common transmission dip at

7975 Hz (dashed line) occurs for both samples, which coincides with the crossing point of the two valley interface state dispersions (orange and purple lines in Figure 1d) at $k_x = 2\pi/3a$.

A quantitative theoretical analysis is pursued after the successful experimental demonstration of acoustic topological interferometers. Knowing the process of the topological interferometer, the transmission coefficient to Port 3 ($T_{13}$) can thus be derived using the scattering matrix method (Supporting Information):

$$T_{12} = \left|\sqrt{\alpha} + \frac{(1-\alpha)\sqrt{1-\alpha}}{\sqrt{\alpha}\sqrt{1-\alpha}+e^{\beta L}e^{-2i(\varphi_1+\varphi_2)}}\right|^2,$$

$$T_{13} = \left|\sqrt{\alpha}e^{-\frac{\beta L}{4}}\left(\sqrt{1-\alpha} - \frac{(1-\alpha)\sqrt{\alpha}}{\sqrt{\alpha}\sqrt{1-\alpha}+e^{\beta L}e^{-2i(\varphi_1+\varphi_2)}}\right)\right|^2, \quad (1)$$

where $\alpha$ is the energy partition rate we obtained for Device 1 in Figure 2a,b. We can thus calculate the transmission spectra with $L = 36a$ (red) and $L = 48a$ (blue) from Eq. (1), also shown in Figure 3b in solid lines. During the propagation along each domain walls of the rhombus-shaped region, a dynamic phase $\varphi_{1(2)} = \Delta k_{1(2)} L/4$ is accumulated. Here, $\Delta k_{1(2)}$ represents the corresponding wave vector difference to $k_x = 2\pi/3a$ among the purple (orange) valley interface states at different frequencies, as shown in Figure 1d. Purple (orange) interface appears on the left and bottom (top and right) domain walls of the interferometer. The attenuation coefficient $\beta = 2.67$, referring to the inevitable sound loss, is obtained by a least square fitting of Eq. (1) to the experimental data to $T_{13}$ at $L = 36a$. From Eq. (1), we can recognize that the spectral oscillation of $T_{13}$ is originated from the phase term on the denominator, where $T_{13}$ shall be minimized and maximized with $2(\varphi_1+\varphi_2) = 2n\pi$ and $2(\varphi_1+\varphi_2) = (2n+1)\pi$, respectively. At 7975 Hz, when the purple and orange interface state dispersions cross at $k_x = 2\pi/3a$, the total dynamic phase accumulation has to be zero, and a transmission dip shall occur at this frequency for all rhombus-shaped interferometers, as observed experimentally.

The existence of π Berry phase has also been pointed out to play a key role in the graphene valley partition interferometers.[18] As shown in the Supporting Information, only by allowing the existence of the π Berry phase accumulating when sound wave transports in a closed loop in our interferometer can reach a consistent result between analytical derivation of Eq. (1) and experiments. For instance, $T_{13}$ becomes maximized at 7975 Hz without including π Berry phase. We thus indirectly experimentally verify the existence of π Berry phase for valley pseudospins in SCs.

In a traditional interferometer, the length difference between two paths provides the interference. From Eq. (1), we find that it is the total circumference that affects the interference effect in the topological interferometer. The backscattering-immune nature of topological interface states makes the shape of the middle domain for interference insignificant.[58] Thus, we can also consider a triangle-shaped interferometer. As only the interface state marked in purple (Figure 1d) exists, different from a rhombus-shaped interferometer, the total dynamic phase accumulation is $\varphi = \Delta k_1 L$. When there exists $\Delta k_1(n) = (2n+1)\pi/L$ ($\Delta k_1(n) = 2n\pi/L$), the $n$th peak (dip) in the transmission spectrum shall appear. The measured transmission spectra for the different circumferences $L = 24a$ (green solid circles) and $L = 36a$ (blue solid circles) are illustrated in Figure 3c, respectively, where spectral oscillations also appear with multiple peaks and dips. We also numerically study larger triangular-shaped interferometers with circumferences $L = 48a$ and $L = 96a$ and record the corresponding transmission peaks and dips to Port 3. Given the interferometer nature, we shall be able to obtain the corresponding wave vector difference $\Delta k_1$ at different frequencies. As shown in Figure 3d, we plot the achieved wave vectors at different frequencies for different circumferences while our numerical and experimental findings are consistent with each other. The nearly linear dispersion obtained agrees with the calculated interface state dispersion (purple solid line in Figure 1d) indicating that we reach an effective approach to precisely measure the topological interface dispersion. The dispersion relation is one of the most important properties of topological states and decides its future applications. The traditional method to measure acoustic dispersion is by a fast Fourier transform (FFT) on the measured pressure field distributions[23,59] and its accuracy largely relies on the spatial precision and the sample size, which is very time-consuming. Now by only measuring the sound wave intensity at exit, this simple method to obtain interface state dispersion can not only largely save the corresponding experimental efforts but also be extended to other systems where direct field mapping is extremely difficult.[16,27,42] By simply analyzing the peaks and dips of the transmission spectrum, a new scheme to measure the valley interface dispersion is provided.

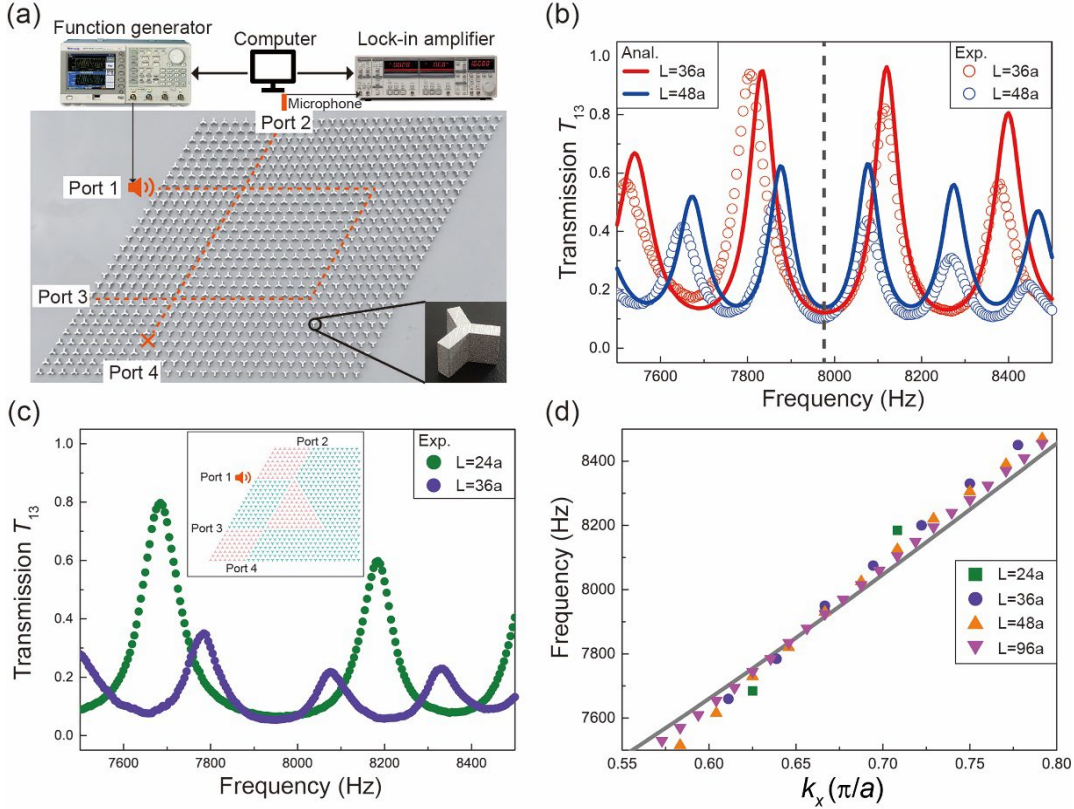

**Figure 3.** a) Photograph of topological valley acoustic interferometer with the experimental set-up. The 3D view of each unit cell is shown in the inset. b) Theoretical (solid line) and measured (circles) transmission spectra to Port 3 for topological interferometers with $L = 36a$ (marked in red) and $L = 48a$ (marked in blue), sharing a common transmission dip at 7975 Hz (grey dashed line). c) Measured transmission spectra to Port 3 for topological interferometers with $L = 24a$ (green solid circles) and $L = 36a$ (purple solid circles). d) Discrete dispersion relation extracted from transmission spectra to Port 3 for triangle-shaped topological interferometers with $L = 24a$ (green solid squares), $L = 36a$ (purple solid circles), $L = 48a$ (sallow solid triangles), and $L = 96a$ (violet solid inverted triangles), compared with numerical dispersion relation (grey solid line).

**4. Tuning Sound Wave Manipulation Based on Valley Acoustic Interferometers**

From Eq. (1), besides the dynamic phase accumulation $\varphi$, various approaches can be used to manipulate sound wave in the interferometer. Though the frequencies of peaks (dips) are governed by topological dispersion relations, the transmission amplitude at each frequency can be effectively tuned by the local geometry design of the valley SC interferometer. As illustrated in Figure 2, different domain configurations can provide arbitrary energy partition rates $\alpha$. Moreover, in Eq. (1) we witness the importance of how $\alpha$ changes the transmission spectra and we can tune the property of our proposed valley acoustic interferometers by tuning $\alpha$. A first straightforward trial will be a geometrical change where we move the locations of Port 3 and Port 4 to tune the sound transmission. In **Figure 4**a,b, we show two configurations of topological interferometer I and II with the same rhombus circumference $L = 36a$ but different locations of Port 3 and Port 4. The corresponding measured transmission spectra to Port 2, Port 3, and Port 4 are shown in Figure 4c,d, respectively. Though the frequencies of the corresponding peaks and dips are unchanged because the same circumference is considered, reassigning the exit SC domain to another position can largely affect the spectra as $\alpha$ considered at the intersection is very different. New analytical equations to represent transmissions $T_{12}$ [Eq. (A9)] and $T_{13}$ [Eq. (A10)] are derived in the Supporting Information and the corresponding results are illustrated along with experimental results with good consistency. Here we provide more examples to verify that with an understanding of the mechanism of valley acoustic interferometer, we can arbitrarily manipulate sound energy partition and a similar mechanism can be extended to other systems.

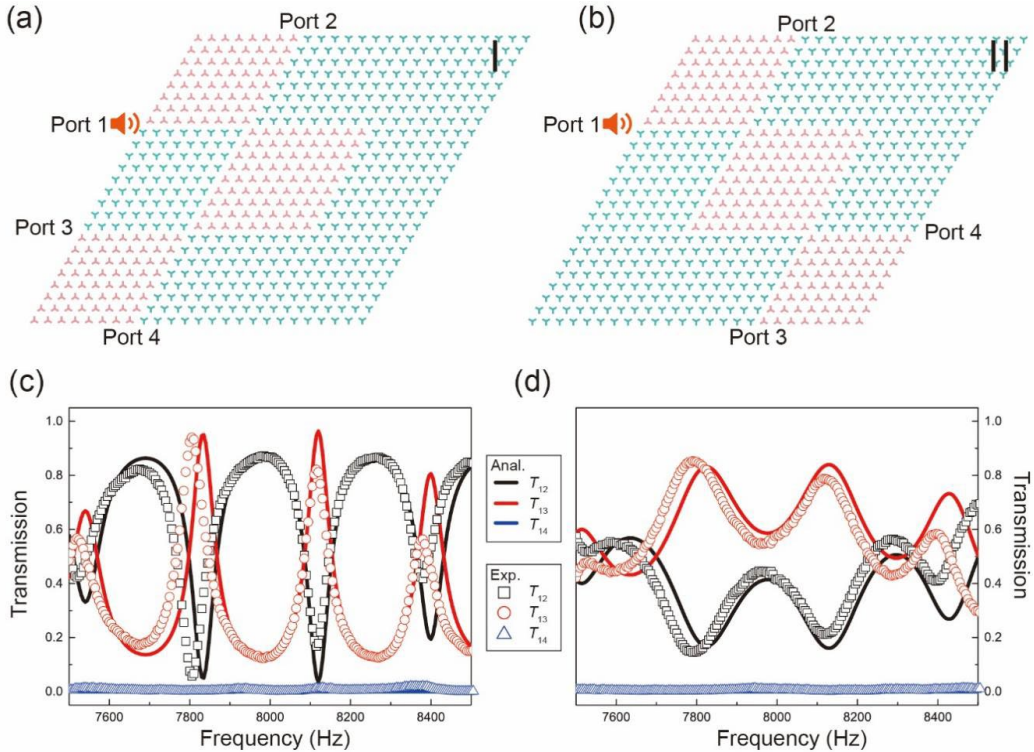

**Figure 4.** Acoustic wave energy manipulation for a fixed circumference topological interferometer with different locations of ports. Schematics of topological interferometer a) I and b) II, whose Port 3 and Port 4 are at the lower-left corner and the lower-right corner, respectively. Theoretical analyses of transmission spectra to Port 2 (black solid lines), Port 3 (red solid lines), and Port 4 (blue solid lines) and measured results of transmission spectra to Port 2 (black open squares), Port 3 (red open circles), and Port 4 (blue open triangles) for interferometer c) I and d) II.

Certain tunability can also be apply to our interferometer design. For instance, by applying rotation to the corresponding aluminum scatterers, changing from topological interferometer I to topological interferometer II is possible, as shown in Figure 4. Moreover, we can also locally change the rotation angles of Y-shape aluminum scatterers around the intersection point (marked in black in **Figure 5**a) to tune the corresponding energy partition ratio $\alpha$ at the intersection may. The transmission to Port 2 and Port 3 with this tunable intersection angle is derived as

$$T_{12} = \left| \sqrt{\alpha_1} + \frac{(1-\alpha_1)\sqrt{1-\alpha_2}}{\sqrt{\alpha_1}\sqrt{1-\alpha_2} + e^{-2i(\varphi_1+\varphi_2)}} \right|^2,$$

$$T_{13} = \left| \sqrt{1-\alpha_1}\sqrt{\alpha_2}\left(1 - \frac{\sqrt{1-\alpha_2}\sqrt{\alpha_1}}{\sqrt{\alpha_1}\sqrt{1-\alpha_2} + e^{-2i(\varphi_1+\varphi_2)}}\right) \right|^2, \quad (2)$$

where $\sqrt{\alpha_1}$ ($\sqrt{\alpha_2}$) delicates the energy partition rate at the first (second) intersection point from Port 1 to Port 2 (from left channel of the interferometer to Port 3). The dynamic phase $\varphi_{1(2)} = \Delta k_{1(2)} L/4$ is also accumulated during the propagation along each domain walls of the rhombus-shaped region. From Eq. (2), energy redistribution between Port 2 and Port 3 can be easily configured with different energy partition ($\sqrt{\alpha_1}$ and $\sqrt{\alpha_2}$) at the two intersections. In Figure 5(b), we plot the transmission $T_{13}$ with varying $\alpha_1$ and $\alpha_2$ values. It is found that the transmission $T_{13}$ can be tuned with a large range from 0 to 1 nearly, indicating that the energy distribution ratio $T_{12}/T_{13}$ can range from 0 to infinity, inferring its infinite potential to manipulate sound wave.

The line shape of the transmission spectrum is another tunable degree of freedom in the interferometer, which can also be asymmetric. For instance, the ultra-sharp line shape of Fano resonance[60] is due to the coupling between a background and a resonant scattering process and can be applied in sensors.[61] However, its design is substantially sensitive to geometrical and environmental changes, which hinders the practical realization of an asymmetric line shape. Here, by coupling with one other topological interferometer, a Fano-like asymmetric line shape is obtained. The schematic of the topological Fano-like resonance interferometer is illustrated in Figure 5c. Two triangle-shaped domains with a slight difference in side length $\Delta L$ (marked in black) are back-to-back configured. With shared in-couple and out-couple domain walls, this system can be considered as a coupling between two interferometers with different sizes. As the spectral characteristic of each interferometer is only circumference dependent, the resonating condition in one interferometer could lead to an antiresonance in the other and their coupling leads to the asymmetric line shape of the transmission spectrum to Port 3. As the simulation results illustrated in Figure 5d, the spectrum exhibits a symmetry line shape when $\Delta L = 0$, while an asymmetric line shape appears when $\Delta L = a$ and $\Delta L = 2a$. Tuning the rotation of valley SCs in certain domains can not only tune the transmission rate as demonstrated in the previous session, but also the line shape of the spectrum can be customized, which enriches and expands the tunability of the interferometer.

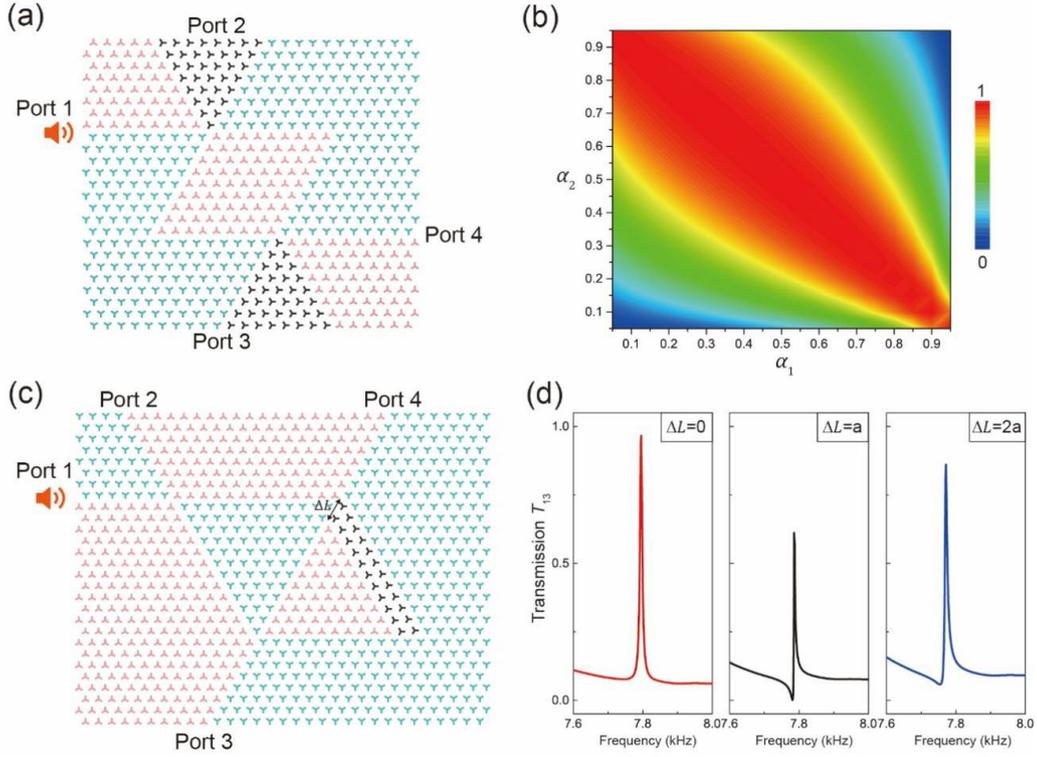

**Figure 5.** a) Schematic of acoustic wave energy manipulation by tuning the rotation angles of Y-shaped aluminum scatterers (marked in black) around the intersection points. b) Analytical transmission to Port 3 when changing the energy partition $\alpha_1$ and $\alpha_2$ at two intersections. c) Schematic of Fano-like resonance interferometer. Two triangle-shaped domains with a slight difference in side length $\Delta L$ (marked in black) are back-to-back configured. d) Simulated transmission spectra to Port 3 with $\Delta L = 0$ (red solid line), $\Delta L = a$ (black solid line), and $\Delta L = 2a$ (blue solid line).

## 5. Conclusion

In summary, we report topological valley acoustic interferometers constructed by Y-shaped aluminum valley SCs. By TB calculation and experimental demonstration, we successfully tune the acoustic energy partition rate by configuring the channel, which is realized by rotating valley SCs. An analytical theory proposed to explain the transmission property of topological valley acoustic interferometers not only matches well with the experimental observations but also verifies that an additional π Berry phase accumulates circling along the shape-independent topological valley acoustic interferometer. Based on the spectral oscillation originating from the accumulated dynamic phase and π Berry phase, a simplified method to measure acoustic valley interface state dispersion is explored, which overcomes the shortcomings of the traditional FFT method and improves the measuring efficiency by simply analyzing the peaks and dips of the measured transmission spectrum. As the frequencies of peaks and dips are determined only by the circumference of the interferometer, an effective approach to tuning its transmissions, as well as the spectral line shapes proposed and realized by the local geometry design of the interferometer, exhibits strong tunability under the same physical mechanism. Therefore, our work opens an avenue to design future acoustic devices with the function of sound wave manipulation based on the physical mechanism of interference and π Berry phase, which may inspire a variety of potential applications and cause deep impacts in various fields.

**Acknowledgements**

W.Z., J.H.C., and S.G.C. contributed equally to this work. This work was supported by the National Natural Science Foundation of China (Grant No. 12274315), the National Key R&D Program of China (Grant No. 2022YFA1404400), and a Project Funded by the Priority Academic Program Development of Jiangsu Higher Education Institutions (PAPD).